\documentclass[aps,pre,twocolumn,showpacs,floatfix]{revtex4} 
\usepackage{amsmath,amssymb,times,bm}
\usepackage[pdftex]{graphicx}

\pdfoutput=1

\newcommand{\be}{\begin{equation}}
\newcommand{\ee}{\end{equation}}
\newcommand{\bea}{\begin{eqnarray}}
\newcommand{\eea}{\end{eqnarray}}
\newcommand{\bi}{\begin{itemize}}
\newcommand{\ei}{\end{itemize}}
\newcommand{\bw}{\begin{widetext}}
\newcommand{\ew}{\end{widetext}}
\newcommand{\kommentar}[1]{}

\newcommand{\rl}{\rangle\langle}

\DeclareMathSymbol{\varTau}{\mathord}{letters}{84}

\begin{document}
 
\title{Directed excitation transfer in vibrating chains by external fields
}
\author{Oliver M{\"u}lken}
\author{Maximilian Bauer}
\affiliation{
Physikalisches Institut, Universit\"at Freiburg,                                                 
Hermann-Herder-Stra{\ss}e 3, 79104 Freiburg, Germany}

\date{\today} 
\begin{abstract} 
We study the coherent dynamics of excitations on vibrating chains. By
applying an external field and matching the field strength with the
oscillation frequency of the chain it is possible to obtain an (average)
transport of an initial Gaussian wave packet. We distinguish between a
uniform oscillation of all nodes of the chain and the chain being in its
lowest eigenmode. Both cases can lead to directed transport.
\end{abstract}
\maketitle

\section{Introduction}

The transport of energy or charge is fundamental for a large variety of
physical, chemical, and biological processes. One of the most prominent
examples is the energy transfer in the light-harvesting complexes in
photosynthesis \cite{fleming2004}. There, the energy of the captured solar
photons is transported via a molecular backbone to the reaction center
where the energy is transformed into chemical energy. Recent experiments
have shown that coherent features of the transport process might be
crucial for a high efficiency  \cite{engel2007,collini2010}. Usually, the
system and the dynamics of the excitation (exciton) is modeled by open
quantum systems where the system of interest, e.g., the light-harvesting
complex, is coupled to an external environment. It has been shown that the
environment can also support the coherent dynamics
\cite{cheng2006,mohseni2008,caruso2009,olaya-castro2008,thorwart2009,ms2010}.

Most of the models assume a time-independent Hamiltonian motivated by the
fact that, indeed, the network of chromophores underlying the energy
transfer is rather static, even at higher (room) temperatures. However,
this need not be the case. One can easily imagine the situation where the
underlying molecule is not static but performs some kind of mechanical
oscillation. Asadian {\sl et al.} have shown that certain types of motions
can enhance the transfer efficiency when compared to the static situation
\cite{asadian2010}. In a related model, Semi{\~a}o {\sl et al.} studied
the modulation of the excitation energies of coupled quantum dots driven by
a nanomechanical resonator mode, also enhancing the transport efficiency
\cite{semiao2010}. Vaziri and Plenio showed that the periodic modulation
of ion channels leads to the emergence of resonances in their transport
efficiency \cite{vaziri2010}.

Another influence on the dynamics can be external fields. Hartmann {\sl et
al.} have shown for the coherent transport of an initial Gaussian wave
packet on a discrete (static) chain of nodes that by suitably switching
the direction of a constant external field, one can achieve directed
transport \cite{hartmann2004}. There, the switching frequency has been
matched with the Bloch oscillation frequency. The effect of Bloch
oscillations on the trapping of excitations has been studied by Vlaming
{\sl et al.}, finding that the trapping efficiency crucially depends on
the strength of the external field (the bias)
\cite{vlaming2007,vlaming2008}.

Clearly, mechanical motions and external fields are not restricted to
energy transfer in molecular aggregates. Other examples include cold atoms
in optical lattices whose spacings can be periodically modulated
\cite{alassam2010} or waveguide arrays where the ``external field'' is
achieved by a linear variation of the effective refractive index across
the array, see, e.g., \cite{christodoulides2003}.

A question we address in this paper is whether it is possible to engineer
the excitation transport in systems performing mechanical oscillations 
with a constant external field such that also here one obtains directed
transport.

\section{Model}

We consider the excitation dynamics on a finite chain of $N$ nodes with
time-dependent couplings $J_n(t)$ between two adjacent nodes of the chain.
The Hamiltonian in the node-basis $\{|n\rangle, n=1..N\}$ reads
\be
\bm H_0 = \sum_{n=1}^N E_n |n\rl n| + \sum_{n=1}^{N-1} J_n(t) \big(
|n\rl n+1| + |n+1 \rl n|\big),
\ee
where the $E_n$ are the site energies.  Now, in addtion we apply an
external field with strength $f$, such that the total Hamiltonian for an
excitation on a vibrating chain reads
\be
\bm H_S = \bm H_0 + f\sum_{n=1}^N n |n\rl n|.
\ee

For chains whose nodes (molecules or atoms) interact via dipole-dipole
forces, the couplings decay with the third power of the distance between
the nodes. For fairly large distances between adjacent nodes the coupling
to next-nearest neighbors can be neglected such that the assumption of
nearest neigbor couplings in $H_0$ can be justified.

Now, there are two competing effects: On the one hand excitations in a
static chain with an external field perform Bloch oscillations
\cite{hartmann2004,holthaus1996,fukuyama1973}. On the other hand the
time-dependent couplings can cause an enhanced transport efficiency
\cite{asadian2010,semiao2010,vaziri2010}. If the oscillations are
periodic, the distances between two adjacent nodes vary in a given
interval. Short distances means stronger couplings and thus faster
transport from node to node. Longer distances lead to weaker couplings and
slower transport. Therefore, matching the Bloch frequency with the
frequency of the chain oscillation should lead to an effective transport
in one direction along the chain. The reason is that in the first half of
the Bloch period $T_B$ the distances between the nodes are smaller while
in the second half of $T_B$ the distances are larger. This leads to
different displacements in the two half periods and consequently to an
overall displacement of the inital excitation in one direction. 

We choose two scenarios for the couplings $J_n(t)$: 

(i) Each node of the chain oscillates uniformly with the same frequency
$\omega$ and with the the same amplitude $a$.  The couplings follow now as 
\be
J_n(t) = J(t) = -V / \Big[1-2a \sin(\omega t + \phi)\Big]^3.
\label{oscillatingJ}
\ee
The same setting has been used by Asadian {\sl et al} \cite{asadian2010}.
In the following we also assume all site-energies to be the same, i.e., we
set $E_n=E=0$.

(ii) The chain is in its lowest eigenmode, such that for the $q$th
eigenmode the couplings $J_{n,q}(t)$ between $n$th and $(n+1)$st node are
\be
J_{n,q}(t) = -V / \Big[1-2a_{n,q} \sin(\omega_q t + \phi)\Big]^3,
\label{eigenmodeJ}
\ee
see Sec.~\ref{eigenmodes} for details.

Clearly, for time-constant $J_n=J$ we recover the known Bloch oscillations
with frequency $\omega_B = f/\hbar$ (we set $\hbar=1$ in the following).
Thus, the period of the oscillation $T_B=2\pi/\omega_B=2\pi/f$.

The dynamics of an initial excitation is governed by the Liouville-von
Neumann equation for the density operator $\bm\rho(t)$. Without any
external environment leading to decoherence, the dynamics is fully
coherent following 
\be
\dot{\bm\rho}(t) = -i [\bm H_S,\bm\rho(t)].
\ee
Now, if the system is coupled to an environment such that the total
Hamiltonian can be split into three parts, $\bm H_{\rm tot} = \bm H_S +
\bm H_R + \bm H_{RS}$, where $H_R$ is the Hamiltonian of the environment
and $H_{RS}$ is the Hamiltonian of the system-environmental coupling. For
small couplings to the environment we will study the dynamics by the
Lindblad quantum master equation \cite{Breuer-Petruccione}
\be
\dot{\bm\rho}(t) = -i [\bm H_S,\bm\rho(t)] -\lambda\sum_{j=1}^N
\big(\bm\rho(t) - \langle j | \bm\rho(t) | j\rangle\big) |j\rl j|,
\label{lqme}
\ee
where we assumed Lindblad operators of the form $\sqrt{\lambda} |j\rl j|$.
The term proportional to $\lambda$ mimicks the influence of the
environment leading to decoherence. In the following we will consider the
occupation probabilities $\rho_{kk}(t) \equiv \langle k | \bm\rho(t) |k
\rangle$ for a given initial condition $\bm\rho(0)$.

\section{Results}

In all calculations shown below we used $N=103$ and an initial Gaussian
wave packet centered at $N_c(0) = N_0$ with a standard deviation of
$\sigma=6$. We adjust $N_0$ such that in the first two periods of the
Bloch oscillations the wave packet does not encounter the edges of the
chain, such that we can exclude interference effects caused by reflection.
We further take $V=1$.

\subsection{Static chain}
\label{sec-statchain}

We start by considering the static chain, i.e., no oscillations ($a=0$).
Without any external field and no external environment, the dynamics of
wave packets on the static chain is very similar to the motion of a
quantum particle in a box \cite{mb2005b,mb2006a}. One can also observe
(partial) revivals of initially localized wave packets caused by
reflections at the end of the chain, thus obtaining the discrete analog of
so-called quantum carpets \cite{kinzel1995,fgrossmann1997}.

\begin{figure}[h]
\centerline{\includegraphics[clip=,width=\columnwidth]{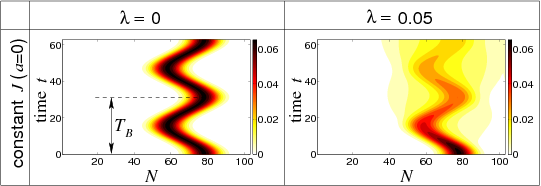}}
\caption{(Color online) Static chain: Contour plot of the occupation
probabilities for $a=0$ and $f=0.2$ with $\lambda=0$ (left panel) and
$\lambda=0.05$ (right panel). Dark (black) regions correspond to large
probabilities, while bright (yellow/white) regions correspond to low
probabilities.}
\label{N103_gauss_bloch}
\end{figure}

When applying an external field, the situation changes.  Figure
\ref{N103_gauss_bloch} shows for $N_0=78$ the well-known Bloch
oscillations in the occupation probabilities $\rho_{kk}(t)$ with Bloch
frequency $\omega_B=f$ for $f=0.2$ with no external coupling, $\lambda=0$,
(left panel) and with small external coupling, $\lambda=0.05$, (right
panel).  One clearly recognizes the oscillation period of $T_B = 2\pi/f =
10\pi$.  The coupling to the environment leads to a spreading of the wave
packet over more and more nodes as time progresses. Eventually, this will
lead to the equilibium distribution.

In a continuous approximation for an infinite line, the position of the
center of the wave packet follows for vanishing initial momentum as
\cite{hartmann2004,holthaus1996,fukuyama1973}
\be
\Delta N (t) \equiv N_c(t) - N_0 \simeq -\frac{4V}{f} \sin^2(ft/2).
\label{deltaNnoJ}
\ee
Obviously, there is no transport after integer values of $T_B$, only after
$T_B/2=\pi/f$ has the wave packet travelled by $|\Delta N (T_B/2)| = 4V/f
= 20$ nodes in the direction of the field. We note that by instantly
reversing the field after $T_B/2$ the wave packet will continue to move to
the left side, such that it is possible to obtain directed transport by
switching the field every half-period, see
\cite{hartmann2004} for details. 

\subsection{Uniformly oscillating chain}

If the chain is not static ($a\neq 0$) but oscillates such that the
couplings are given by Eq.~(\ref{oscillatingJ}), it is possible to obtain
- on average - a net transport of the wave packet in one direction.
However, this will depend on the choice of the field strength $f$, i.e.,
on the frequency of the Bloch oscillation, on the phase shift $\phi$, and
on the amplitude $a$.

\subsubsection{Analytical approximation}

Before turning to the numerical results, we give an analytical estimate of
the displacements $\Delta N_l \equiv  [N_c(l T_B) - N_0]$ ($l\in\mathbb
N$) of the center of the wave packet after integer values of the Bloch
period $T_B$.  For the static (infinite) chain, starting from
Eq.~(\ref{deltaNnoJ}) and differentiating with respect to time, one has
\be
\dot N(t) = -4V \sin(ft/2) \cos(ft/2) = -2V \sin(ft),
\ee
which gives the temporal change of the displacement. Thus, the rate of
transport from node to node is $V$.  We extend this idea to the
oscillating chain and replace the coupling $V$ with the time-dependent
coupling $J(t)$. Then, we define the approximate displacements by
integrating $\dot N(t)$ over integer values of $T_B$:
\be
\Delta N_{l,{\rm approx}} \equiv 
-2V \int\limits_0^{l T_B} dt \ \frac{\sin(ft)}{\big[1-2a\sin(\omega
t+\phi)\big]^3}.
\label{approxN}
\ee
For $\omega=f$ this leads to 
\be
\Delta N_{l,{\rm approx}}  =
-\frac{12l\pi
aV\cos\phi}{f(1-4a^2)^{5/2}}.
\ee
Clearly, the displacement is maximal for $\phi=0$ and minimal (zero) for
$\phi=\pi$. Note that $\Delta N_{l,{\rm approx}}$ is only valid for the
infinite chain.  In the following we will compare $\Delta N_{l,{\rm
approx}}$ to numerical results obtained from Eq.~(\ref{lqme}). As we will
show, for the uniformly oscillating chain, $\Delta N_{l,{\rm approx}}$
agrees very well with the numerical results. Also for the chain in its
lowest eigenmode we will use $\Delta N_{l,{\rm approx}}$ as a starting
point to define an adhoc fitting function $\Delta N_{l,{\rm fit}}$ which
also turns out to be in very good agreement with the numerical results.

\subsubsection{Numerical results}

\begin{figure}[t]
\centerline{\includegraphics[clip=,width=\columnwidth]{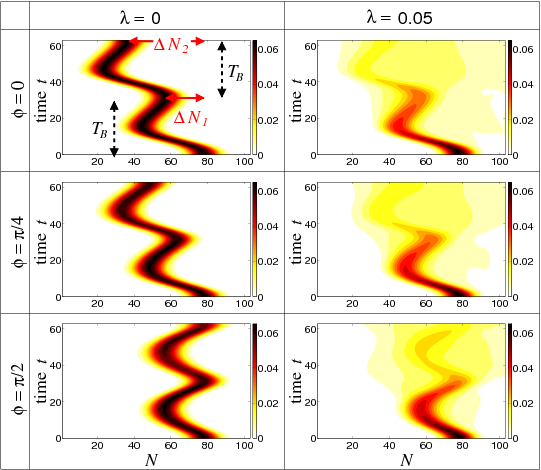}}
\caption{(Color online) Oscillating chain:  Contour plot of the occupation
probabilities $\rho_{kk}(t)$ for $a=0.1$ and $\omega=f=0.2$ with
$\lambda=0$ (left panels) and $\lambda=0.05$ (right panels). The three
rows correspond to different values of $\phi=0,\pi/4$ and $\pi/2$,
respectively.}
\label{N103_gauss}
\end{figure}

Figure \ref{N103_gauss} shows the occupation probabilities $\rho_{kk}(t)$
for the case $\omega_B = f = \omega=0.2$ with $a=0.1$ and for different
phase shifts $\phi$. Again, the left panels show the results for isolated
chains ($\lambda=0$) and the right panels for small couplings to an
external environment ($\lambda=0.05$). Plots in different rows correspond
to different $\phi$. Matching $f$ with $\omega$ and having no phase shift
results - on average - in a directed transport of the initial wave packet
in the direction of the field.  In the second half of each Bloch period
$T_B$ the wave packet moves in the opposite direction.  However, this is
overcompensated by the motion in the direction of the field in the first
half of each period.

\begin{figure}[b]
\centerline{\includegraphics[clip=,width=\columnwidth]{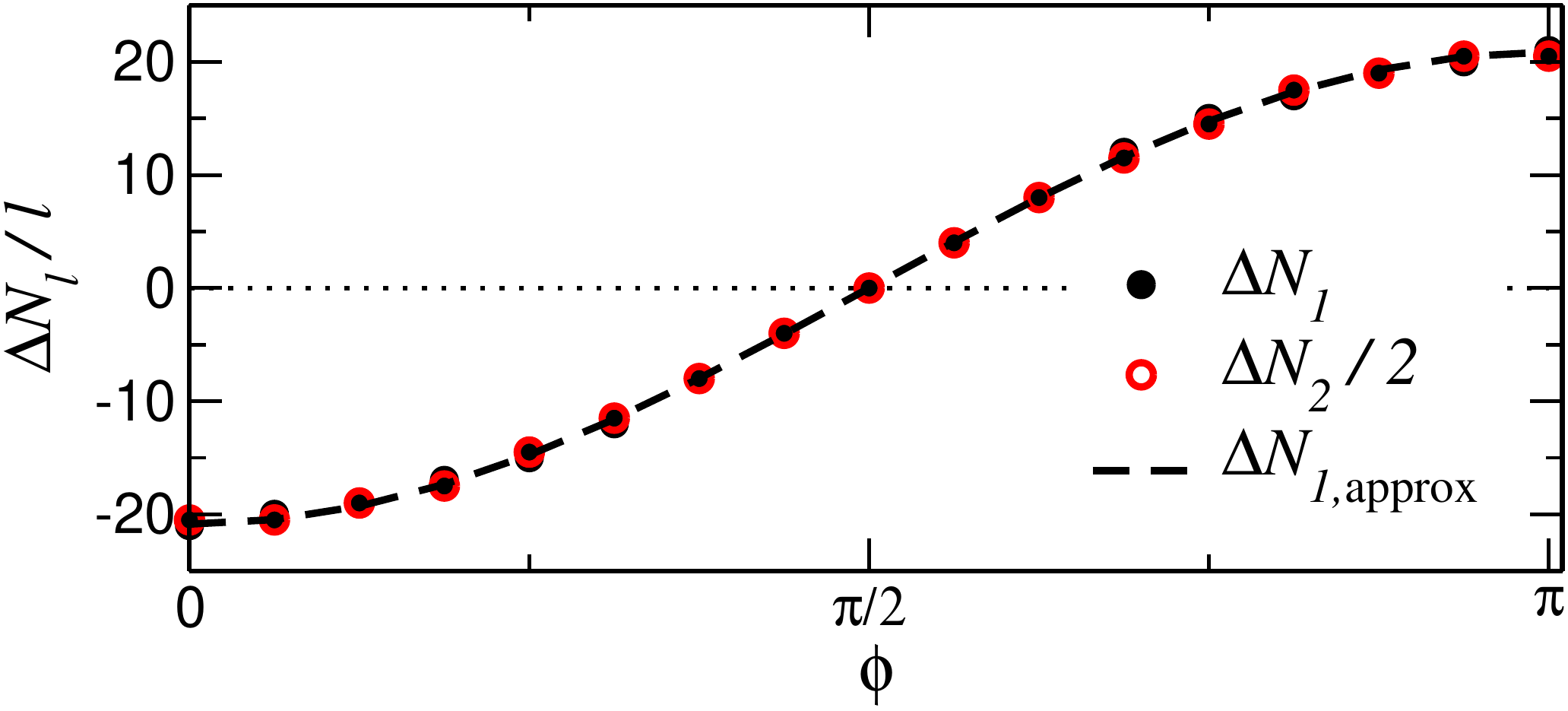}}
\caption{(Color online) Oscillating chain: Displacements $\Delta N_l/l$
with $l=1,2$, extracted from the occupation probabilities for $N_0=52$, as
a function of $\phi$ for $a=0.1$, $\omega=f=0.2$ and $\lambda=0$.  The
dashed lines show $\Delta N_{l,{\rm approx}}$ given in
Eq.~(\ref{approxN}).}
\label{alphaphi}
\end{figure}

The dependence on the phase shift can be expressed by only considering the
average displacement $\Delta N_l$.  Figure \ref{alphaphi} shows the
dependence of $\Delta N_1$ and $\Delta N_2/2$ on $\phi$ for the same
parameters as in Fig.~\ref{N103_gauss}, but with $N_0=52$.  Changing the
initial condition to the center of the chain allows to vary $\phi$ in
between $0$ and $2\pi$ and thus avoiding interference effects due to
reflections at the ends of the chain. Note that this has no influence on
the dynamics because the couplings in the chain are translational
invariant. We distinguish between $\Delta N_1$ after one and $\Delta N_2$
after two periods because, in general, one cannot expect a linear behavior
of $\Delta N_l$ in $l$. However, as it turns out $\Delta N_l$ is
approximately linear in $l$ for the uniformly oscillating chain.

Changing the phase shift allows to control the transport: No phase shift
($\phi=0$) results in values of $\Delta N_1 \approx - 21$ after one
period. A phase shift of $\phi=\pi/2$ results in a behavior similar to the
Bloch oscillations in the static chain, i.e., no transport, see also
Fig.~\ref{N103_gauss_bloch}. Increasing $\phi$ further leads to a reversed
motion, i.e., the wave packet moves ``uphill'' against the direction of
the field. For $\phi=\pi$ the maximal displacement after one period of
$\Delta N_1 \approx 21$ is obtained. For the uniformly oscillating chain,
the values for $\Delta N_2/2$ coincide with the ones for $\Delta N_1$
leading to the linear behavior $\Delta N_l = l \Delta N_1$. In addition,
Fig.~\ref{alphaphi} shows the analytical estimate of Eq.~(\ref{approxN})
which agrees with the numerical results.

\begin{figure}[t]
\centerline{\includegraphics[clip=,width=\columnwidth]{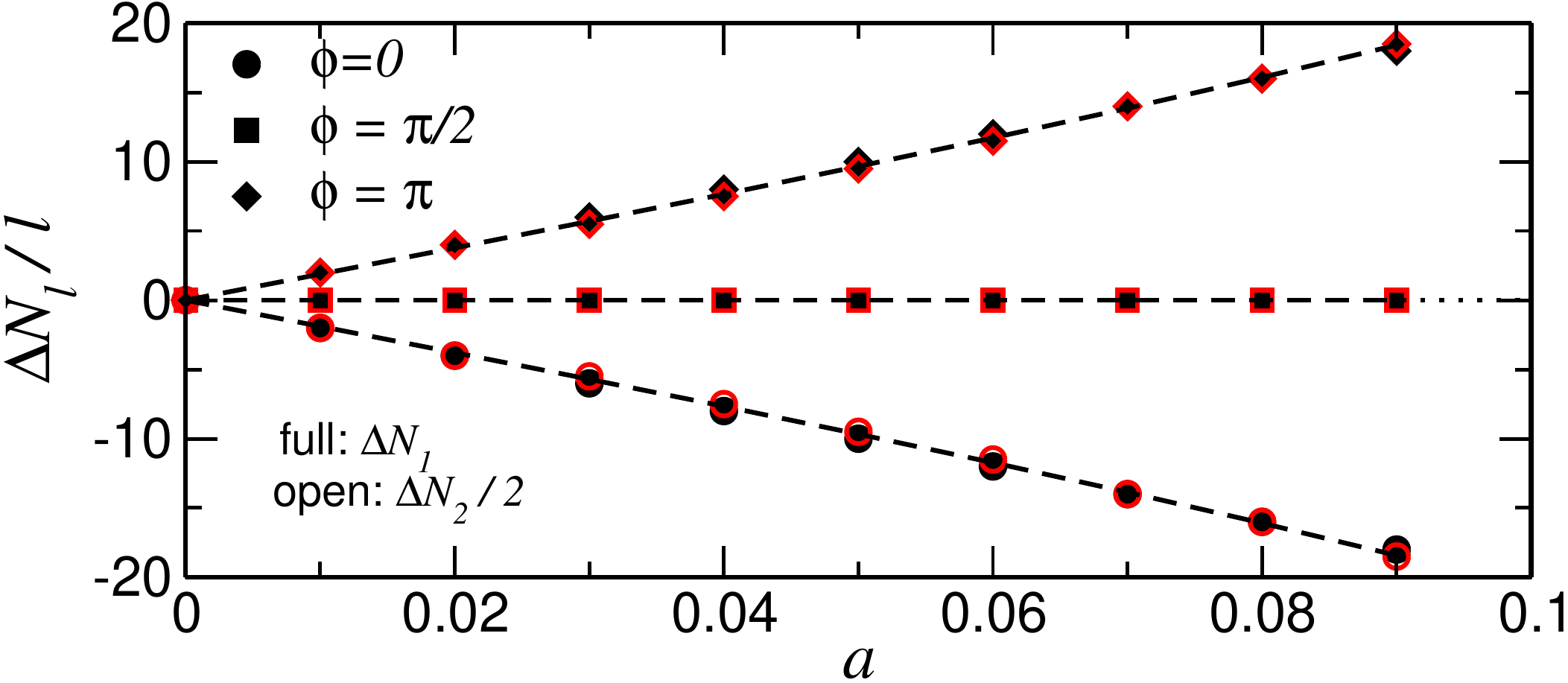}}
\caption{(Color online) Oscillating chain: Displacements $\Delta N_l/l$
with $l=1,2$ as a function of $a$ for $N_0=52$, $\omega=f=0.2$,
$\phi=0,\pi/2,\pi$ and $\lambda=0$. The dashed lines show $\Delta
N_{l,{\rm approx}}$ given in Eq.~(\ref{approxN}).}
\label{alphaa}
\end{figure}
The magnitude of the displacements $\Delta N_l$ also depends on $a$.
Figure~\ref{alphaa} shows $\Delta N_l/l$ as a function of $a$  for
$N_0=52$ and $\phi=0,\pi/2$, and $\pi$. While for $\phi=\pi/2$ there is no
displacement after integer values of $T_B$, the displacements for $\phi=0$
and for $\phi=\pi$ grow with increasing $a$. Again, the dashed lines show
the approximation $\Delta N_{1,{\rm approx}}$ which nicely agrees with the
numerical results.

\begin{figure}[b]
\centerline{\includegraphics[clip=,width=\columnwidth]{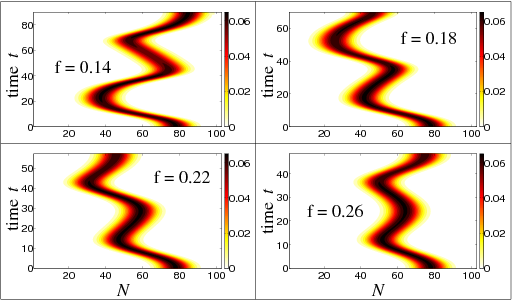}}
\caption{(Color online) Oscillating chain: Contour plot of the occupation
probabilities $\rho_{kk}(t)$ with $\omega=0.2$, $a=0.1$, $\phi=0$, and
$\lambda=0$ for different values of $f$.}
\label{N103_detuned_gauss}
\end{figure}
The effect of having directed transport depends on having the field
strength in resonance with the chain oscillation frequency. In order to
see how crucial the exact matching of $f$ and $\omega$ is, we study
slightly detuned frequencies $\omega_B$, i.e., a mismatch between $\omega$
and $f$. Figure~\ref{N103_detuned_gauss} shows for $\phi=0$ (leading to
maximal $\Delta N_l$ for $\omega=f$) and for $\omega=0.2$ the occupation
probabilities $\rho_{kk}(t)$ for different values of $f$.  Note that
changing $f$ also changes the Bloch period $T_B=2\pi/f$, thus the time
axes are different for different $f$.  A field strength of $f=0.18$ or
$f=0.22$ reflects a detuning by $\pm10\%$ of $\omega$. This still results
in an average directed transport after two periods of $\Delta N_2 \approx
30$ for $f=0.18$ and of $\Delta N_2 \approx 36$ for $f=0.22$. Increasing
the detuning further diminishes the transport. 

\begin{figure}[t]
\centerline{\includegraphics[clip=,width=\columnwidth]{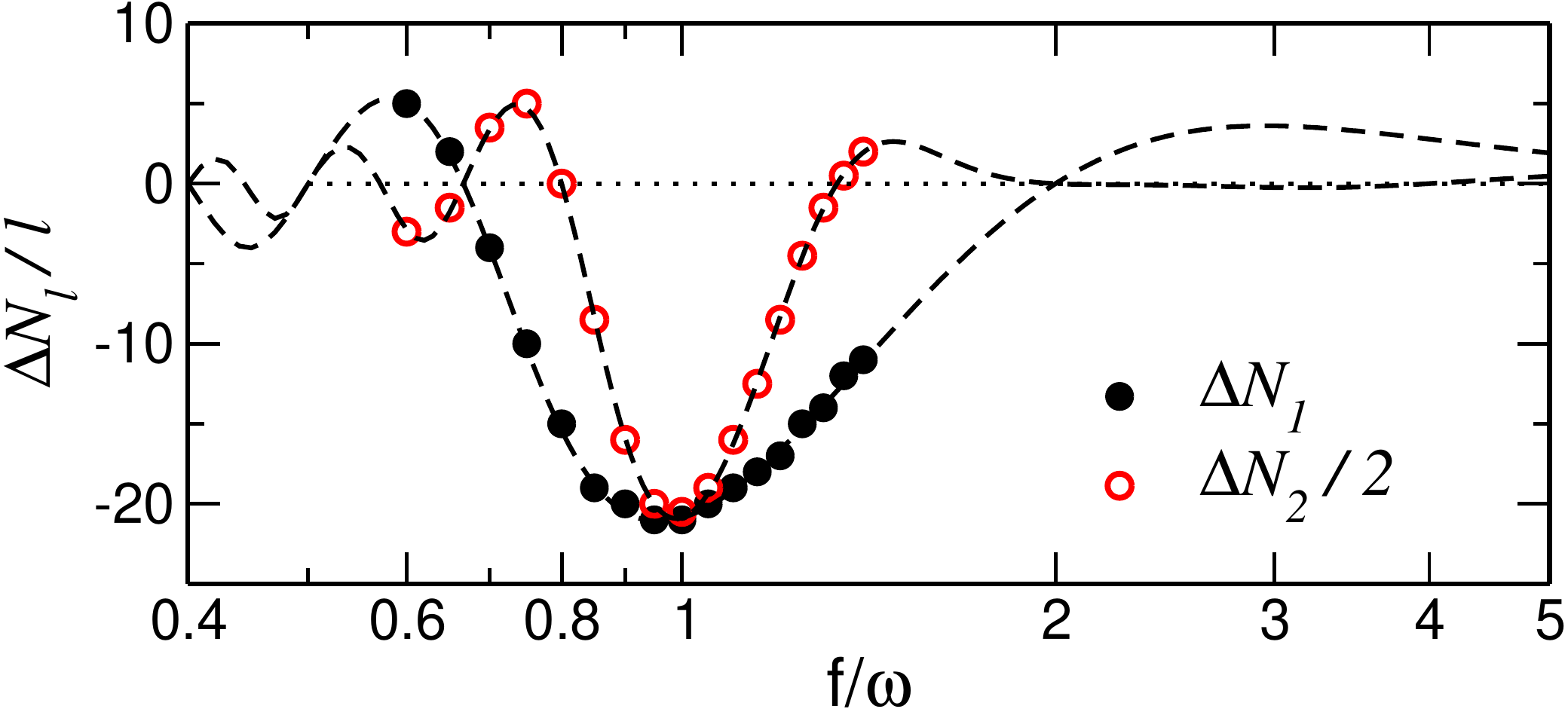}}
\caption{(Color online) Oscillating chain: Displacements $\Delta N_l/l$
with $l=1,2$ as a function of $f$ for $a=0.1$, $\omega=0.2$, $\phi=0$, and
$\lambda=0$ (note the semi-logarithmic scale). The dashed lines show the
approximation obtained by numerical integration, see text for details.}
\label{alphaf}
\end{figure}
Figure~\ref{alphaf} shows the displacements $\Delta N_l/l$ for $\phi=0$
and different values of $f$.  The maximal displacement is obtained for
$\omega\approx f$, as expected.  Decreasing or increasing $f$ results in
smaller displacements: For $f>\omega$ the decrease in displacement is
slower than for $f<\omega$. One also observes that the displacements
change directions. For $f<\omega$, $\Delta N_2/2$ changes direction at
about $f/\omega=0.8$ and $\Delta N_1$ at about $f/\omega\approx 0.67$. For
$f>\omega$, the direction change happens at larger deviations from the
resonace condition. Additionally, there are maximal displacements in the
opposite direction. 

As before, we can obtain an approximation to the numerical results:
Considering now $f\neq\omega$ in Eq.~(\ref{approxN}) and numerically
integrating over integer values of the Bloch oscillation yields the dashed
curves shown in Fig.~\ref{alphaf}. Again, the approximation is in very
good agreement with the numerical data.

Having now explored a large region of the parameters $f/\omega$, $a$, and
$\phi$, we see that the dynamics of an initial Gaussian wave packet can be
manipulated by a suitable choice of these parameters: We can make the wave
packet move - on average - in one preferred direction by choosing the
phase shift $\phi$. The magnitude of the displacements in either direction
is given by $a$. Moreover, we do not have to exactly match the Bloch
frequency $\omega_B=f$ with the oscillation frequency $\omega$ in order to
obtain directed transport, there is a fairly large range of roughly
$\pm10\%$ around $f/\omega=1$ in which large displacements can be
obtained.

\subsection{Chain in lowest eigenmode}
\label{eigenmodes}

In contrast to the previous section, we now consider the dynamics on a
finite chain in its lowest eigenmode. Although this mode is similar to the
uniform oscillation, the finite size of the chain becomes crucial leading
to a non-uniform oscillation of the nodes.

The couplings $J_{n,q}(t)$ in Eq.~(\ref{eigenmodeJ}) between the nodes are
obtained from a normal mode analysis of a free chain of nodes connected by
springs, see \cite{asadian2010} for details. Although the motion of the
nodes is not uniform \cite{rosenstock1955}, there are close similarities
to the results presented in the previous section. 

In order to obtain comparable results we have to adjust the amplitudes and
frequencies according to the couplings  $J_{n,q}(t)$ between nodes $n$ and
$n+1$ for the $q$th eigenmode. The couplings in Eq.~(\ref{eigenmodeJ}) can
be written as
\bea
J_{n,q} (t) &=& -V \Bigg[ 1 
-\frac{2a\sin[2\omega t \sin(q\pi/2N) +
\phi]}{\cos(q\pi/2N)} 
\nonumber \\
&& 
\times \sin(nq\pi/N) \sin(q\pi/2N)
\Bigg]^{-3},
\eea
such that one has
\be
a_{n,q} \equiv \frac{a \sin(nq\pi/N)\sin(q\pi/2N)}{\cos(q\pi/2N)}
\label{ampl_eigenmode}
\ee
and
\be
\omega_q \equiv 2\omega \sin(q\pi/2N).
\ee
Thus, in the following we will use $\omega_q=f$ as the resonance condition
for the frequency and the field. For the amplitude $a_{n,q}$ to be
comparable to the amplitudes in the previous section, we consider the
average absolute value of the amplitudes, i.e.,
\bea
\bar a_q &\equiv& \frac{1}{N}\sum_{n=1}^N |a_{n,q}| =
\frac{a}{N}\tan(q\pi/2N) \sum_{n=1}^N |\sin(nq\pi/N)|
\nonumber \\
&=& \frac{aq}{N} \tan(q\pi/2N) \cot(q\pi/2N) = \frac{aq}{N}.
\eea
Thus, we consider amplitudes $\bar a_q$ which - on average - are of the
same order as the ones in the previous section. This means that we choose
the parameter $a$ in Eq.~(\ref{ampl_eigenmode}) to be $a= N \bar a_q /q$.

\begin{figure}[b]
\centerline{\includegraphics[clip=,width=\columnwidth]{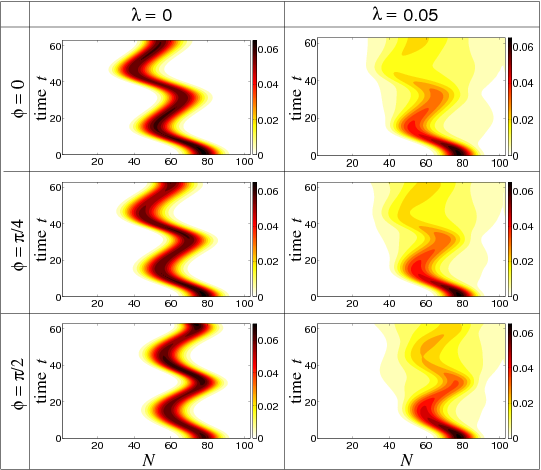}}
\caption{(Color online) Lowest eigenmode: Contour plot of the occupation
probabilities $\rho_{kk}(t)$ for $\bar a_1=0.04$ and $\omega_1=f=0.2$ with
$\lambda=0$ (left panels) and $\lambda=0.05$ (right panels). The three
rows correspond to different values of $\phi=0,\pi/4$ and $\pi/2$,
respectively.  }
\label{N103_q1_gauss}
\end{figure}

Similarly to Fig.~\ref{N103_gauss}, Fig.~\ref{N103_q1_gauss} shows the
occupation probabilities $\rho_{kk}(t)$ for the case
$\omega_B=f=\omega_1$. All plots in Fig.~\ref{N103_q1_gauss} show results
for $\bar a_{1}=0.04$. We use $\bar a_{1}=0.04$ because this clearly
avoids interference effects due to reflections at the ends of the chain.
Coupling this system to an external environment leads, again, to
decoherence and a spreading of the initial wave packet.

\begin{figure}[t]
\centerline{\includegraphics[clip=,width=\columnwidth]{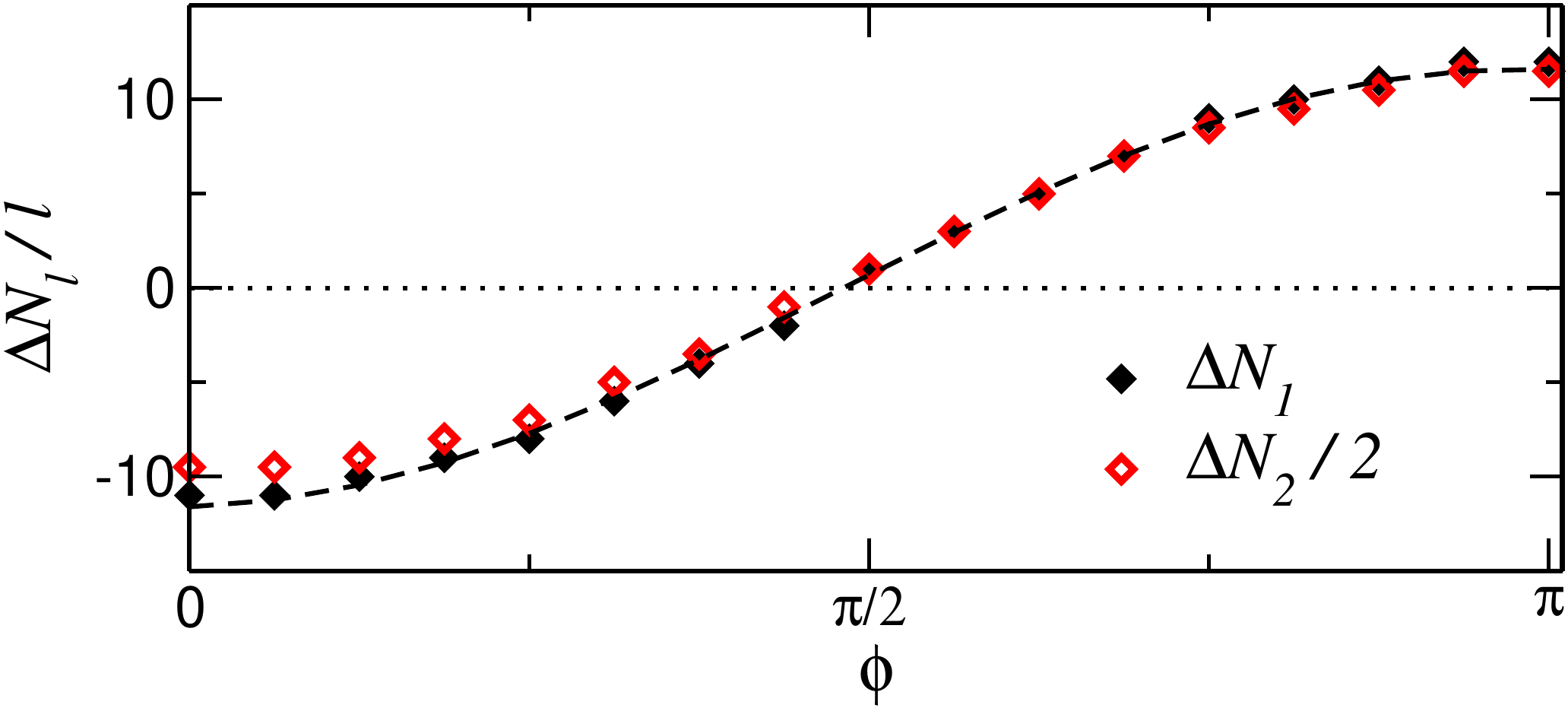}}
\centerline{\includegraphics[clip=,width=\columnwidth]{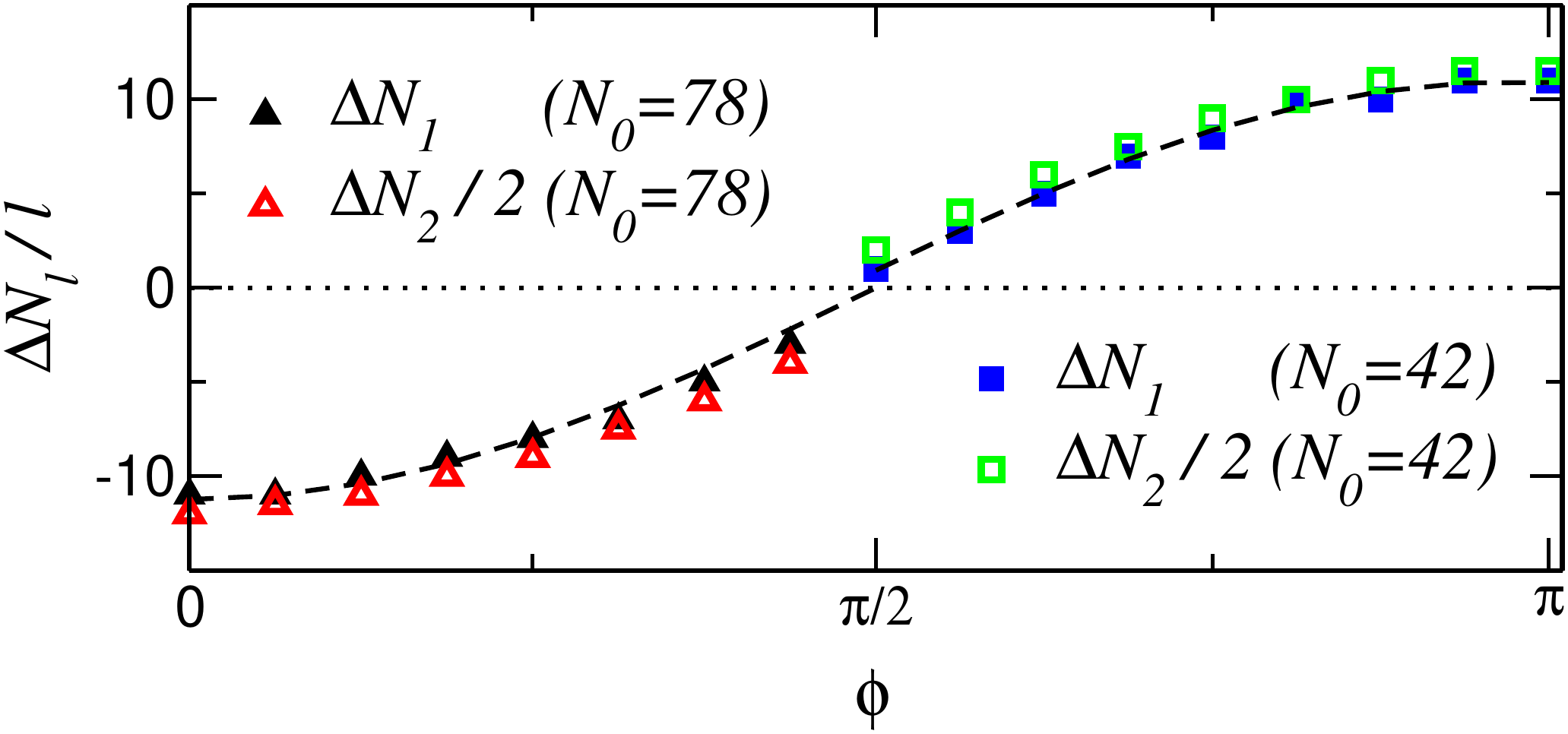}}
\caption{(Color online) Lowest eigenmode: Displacements $\Delta N_l/l$
with $l=1,2$ for $N_0=52$ (upper panel) and for $N_0=42$ and $N_0=78$
(lower panel) as a function of $\phi$ for $\bar a_1=0.04$,
$\omega_1=f=0.2$, and $\lambda=0$. The dashed lines show the fits for
$\Delta N_{l,{\rm fit}}$ given by Eq.~(\ref{deltaNfit}).}
\label{alphaphieigen}
\end{figure}

Figure~\ref{alphaphieigen} shows a comparison of the displacements $\Delta
N_l$ as a function of the phase shift $\phi$ for different $N_0$. Already
for the central initial node, $N_0=52$ (upper panel), one notices the
asymmetry between the behavior of $\Delta N_1$ and $\Delta N_2/2$ for
values of $\phi\in[0,\pi/2]$ and values of $\phi\in[\pi/2,\pi]$. For
$\phi>\pi$ the difference between $\Delta N_1$ and $\Delta N_2/2$ is
smaller than for $\phi<\pi/2$, see in particular the points for $\phi=0$ and
$\phi=\pi$. One also notices that $\phi=\pi/2$ yields $\Delta N_l\neq0$, in
contrast to the uniformly oscillating chain.  However, the overall
behaviors for the two chains are very similar.  Therefore, we fit our
numerical result for $\Delta N_1$ by a cosine, as suggested by
Eq.~(\ref{approxN}), namely, we use 
\be
\Delta N_{l,{\rm fit}} \equiv
\beta_l\frac{\bar a_1\cos(\phi+\alpha_l)}{(1-4\bar a_1^2)^{5/2}},
\label{deltaNfit}
\ee
where $\alpha_l$ and $\beta_l$ are ($l$-dependent) fit parameters. This
already yields a very good agreement with the numerical results, see the
dashed lines in Fig.~\ref{alphaphieigen}.
 
Changing the initial node $N_0$ influences the behavior of the wave
packet. Figure~\ref{alphaphieigen} also shows the behavior of $\Delta N_1$
and $\Delta N_2/2$ for $N_0=42$ (lower panel, right half) and $N_0=78$
(lower panel, left half). While for $N_0=52$ one has $|\Delta N_1|\geq|\Delta
N_2/2|$, one observes for $N=42$ and for $N_0=78$ that $|\Delta
N_1|\leq|\Delta N_2/2|$. However, for all initial nodes shown in
Fig.~\ref{alphaphieigen}, the maximal displacements (for $\phi=0$ and
$\phi=\pi$) are in the same region about $|\Delta N_l/l| \approx 12$.

The slight asymmetry can be attributed to the non-uniform, i.e.,
non-translational invariant, motion of the nodes of the chain and the
additional influence of the external field, which breaks the point
symmetry with respect to the center.

\begin{figure}[h]
\centerline{\includegraphics[clip=,width=\columnwidth]{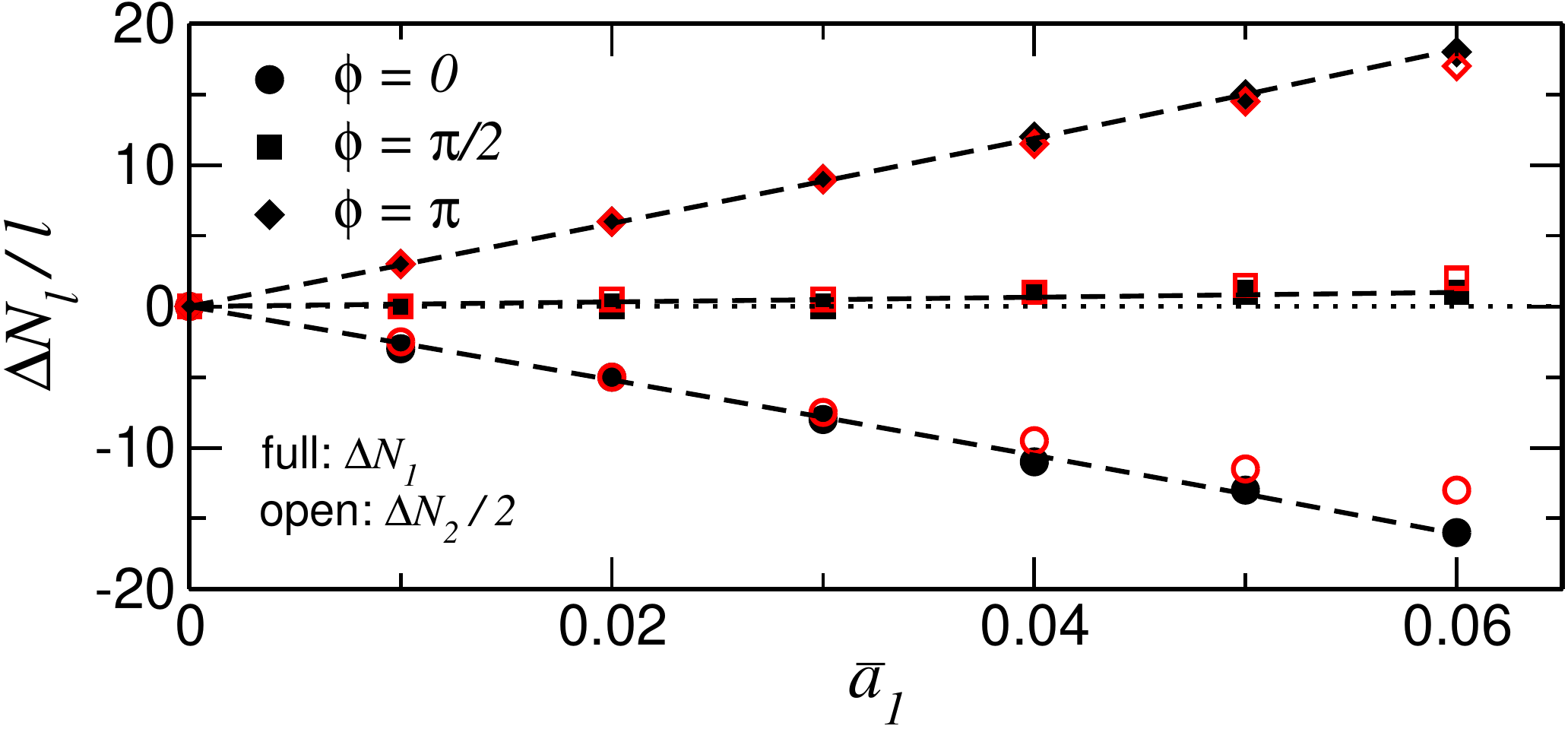}}
\centerline{\includegraphics[clip=,width=\columnwidth]{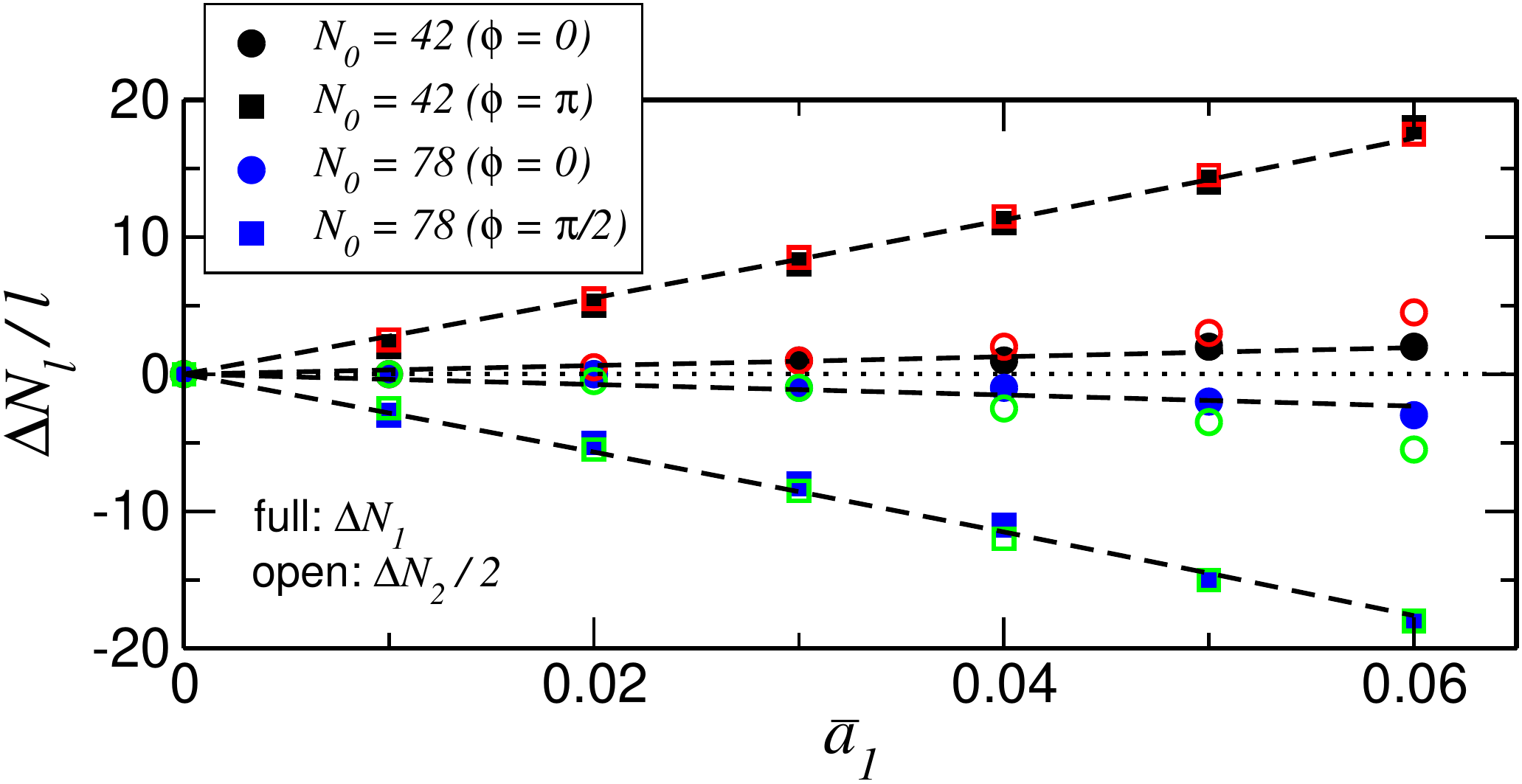}}
\caption{(Color online) Lowest eigenmode: Displacements $\Delta N_l/l$
with $l=1,2$ as a function of $\bar a_1$ for $\omega_1=f=0.2$, and
$\lambda=0$: upper panel for $N_0=52$ with $\phi=0,\pi/2, \pi$ and lower
panel for $N_0=42$ with $\phi=\pi/2, \pi$ and $N_0=78$ with $\phi=0,
\pi/2$. The dashed lines show the fits for $\Delta N_{1,{\rm fit}}$ given
by Eq.~(\ref{deltaNfit}).}
\label{alphaaeigen}
\end{figure}

The $\bar a_1$-dependence of the displacements is shown in
Fig.~\ref{alphaaeigen}. Although the absolute values of $\Delta N_l/l$ are
different for different $N_0$, there is a similar behavior for different
values of $\phi$. Moreover, the behavior is similar to the one for the
uniformly oscillating chain, see Fig.~\ref{alphaa}. Therefore, we fit the
$\bar a_1$-dependence of $\Delta N_1$ by  $\Delta N_{1,{\rm fit}}$ given
in Eq.~(\ref{deltaNfit}). Also here are the fits in very good agreement
with the numerical results.

\begin{figure}[h]
\centerline{\includegraphics[clip=,width=\columnwidth]{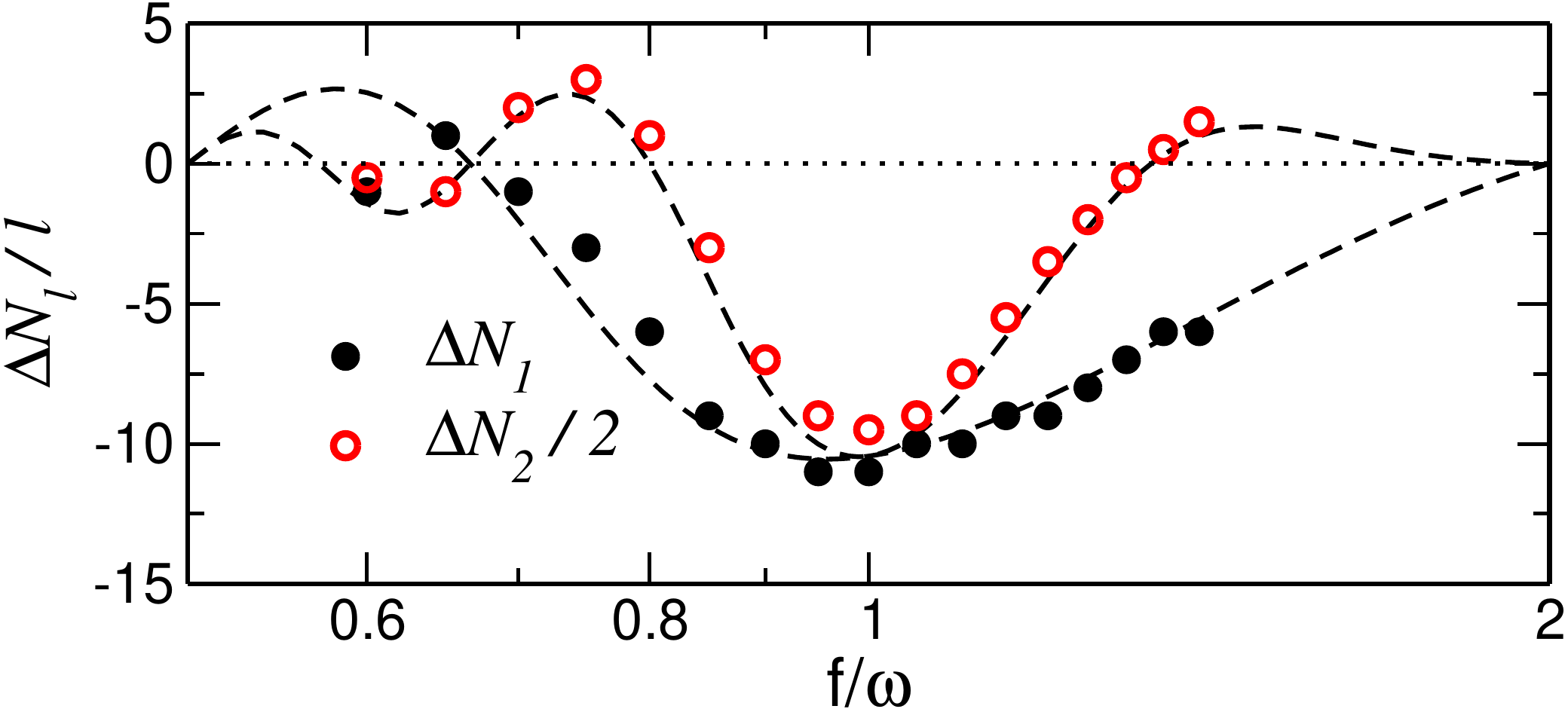}}
\caption{(Color online) Lowest eigenmode: Displacements $\Delta N_l/l$
with $l=1,2$ as a function of $f$ for $\bar a_1=0.04$, $\omega_1=0.2$,
$\phi=0$, and $\lambda=0$ (note the semi-logarithmic scale). The dashed
lines show the approximations of Fig.~\ref{alphaf} scaled by a factor of
$1/2$, see text for details.}
\label{alphafeig}
\end{figure}

Figure~\ref{alphafeig} shows the displacements $\Delta N_1$ and $\Delta
N_2/2$ as a function of $f$ for $\phi=0$. Similar to the oscillating
chain, the displacements are maximal for $f\approx \omega_1$. The dashed
lines show the approximations obtained for the oscillating chain (see
Fig.~\ref{alphaf}) but rescaled by a factor $1/2$. Already this rough
approximation yields good agreement to the numerical results. However, the
points for $f/\omega=0.6$ have to be considered with care, because such a
detuning leads to interference effects due to reflection at the end node
of the chain after one half period. This interference obviously can
influence the dynamics of the wave packet.

Now, also for the chain in its lowest eigenmode we obtain similar results
to the ones for the oscillating chain. However, the absolute values of the
parameters are different. Nevertheless, the approximations given by
Eq.~(\ref{approxN}) turn out to give qualitatively the correct behavior.
Therefore, the same conclusions as for the oscillating chain apply here.

\section{Conclusions}

We have studied the coherent transport of excitations on a finite chain
with time-dependent couplings between adjacent nodes of the chain and in
the presence of an external field. The field leads to Bloch oscillations
while regular time-dependent couplings can lead to an increased transport
efficiency of excitations along the chain.  We showed for uniformly
oscillating chains and for a chain in its lowest eigenmode that matching
the Bloch oscillation frequency with the frequency of the chain leads to
an (average) directed displacement of an initial Gaussian wave packet.
Applying a phase difference allows to manipulate the direction of the
transport, while changing the amplitude of the regular oscillation allows
to manipulate the strength of the displacements.  We corroborate our
findings by an analytic (continuous) approximation for the average
displacement of an initial Gaussian wave packet in an infinite chain after
integer values of the Bloch period. For the uniformly oscillating chain,
this ansatz yields a functional form for the displacements, which agrees
very well with the numerical data. Using the same functional form also
allows to define a fitting function for the chain in its lowest eigenmode,
also leading to very good agreement with the numerical results. In both
cases, interference effects due to reflections at the ends of the chains
have been neglected. 

\begin{acknowledgments}
Support from the Deutsche Forschungsgemeinschaft (DFG) is gratefully acknowledged.
We thank Alexander Blumen for continuous support and fruitful discussions.
\end{acknowledgments}


\end{document}